\documentclass{article}
\usepackage[utf8]{inputenc}
\usepackage{amssymb}
\usepackage{amsmath}
\usepackage{svg}
\usepackage{graphicx}

\title{Smart Contracts- Vulnerabilities, CodeLlama Usage and Gas driven Detection  }
\author{Natan Katz}
\date{December 2023}

\vspace{0.9cm}

\usepackage{csquotes}
\usepackage[utf8]{inputenc}
\usepackage[english]{babel}
\usepackage{subcaption}
\usepackage{graphicx}
\usepackage{listings}
\usepackage{nameref} 
\graphicspath{{plot_fold/}}
\providecommand{\keywords}[1]
{
  \small	
  \textbf{\textit{Keywords--}} #1
}

\begin{document}
\maketitle
 
\begin{abstract}
Smart contracts are a major tool in Ethereum transactions. Therefore hackers can exploit them by adding code vulnerabilities to their sources and using these vulnerabilities for performing malicious transactions. 
This paper presents two successful approaches for detecting malicious contracts: one uses opcode and relies on GPT2 and the other uses the Solidity source and a LORA fine-tuned CodeLlama.  Finally, we present an XGBOOST model that combines gas properties and Hexa-decimal signatures for detecting malicious transactions. This approach relies on early assumptions that maliciousness is manifested by the uncommon usage of the contracts' functions and the effort to pursue the transaction.
\end{abstract}

\keywords{LLM, Fine-tuning, CodeLlama, Smart contracts, Ethereum, ABI, Bayesian networks, Deep Learning, Causal Inference}

\section{Introduction}
Machine learning \textbf{ML} has become a tool in various technological domains. The appearance of tools such as \textbf{ChatGPT} \cite{chat} and \textbf{DALL-E} \cite{dall} made ML accessible to an enormous number of users. However, ML learning influences even more aspects than the common generative approaches: we can see the usage of ML in the medical world, in financial and even in less quantitative disciplines such as archaeology and history. It intrigues nearly everyone that when we provide data samples to software, we enable it to extract novel insights. In parallel to the rise of ML, cyber security has gained more focus lately. The increase in network traffic, particularly in money transactions and the usage of financial information, made most of us vulnerable to malicious web activities. One may expect that cyber and ML will constantly shake hands.  Indeed, we see a massive presence of ML in cyber-security applications. However, creating an ML solution to cyber is often a challenging task due to the following reasons:
\begin{itemize}
\item \textbf{Imbalanced data}- Most of the traffic in cyber is benign: we seldom have traffic that contains more than a single percent of malicious data. Since imbalanced data is one of the classical obstacles in ML,  we may suffer from a tall hurdle.
\item \textbf{Temporality}- In most of the common ML applications, such as image detection or sentiment analysis \cite{sentim}, we can confidently assume that we sample the data from a fixed distribution. In cyber security, the data is often temporal since the attacker and the defender experience an evolution.
\item \textbf{Labeling}- Unlike vision or text, where it is easy to label the data in advance (every junior high pupil can say if the image is of a dog or a cat), labeling cyber data requires expert knowledge.
\end{itemize}
These obstacles impose an effort in nearly every cyber ML project. 
Performing ML projects on blockchain may require even more effort since it may present novel challenges.  This paper describes potential cyber threats in the Blockchain and ML methods to solve them.

\section{Smart Contracts}
\label{smart0}
The following section introduces the concept of smart contracts and the potential problems we may encounter when using them that they may rise.

\subsection{What is Ethereum?}
In 2013, Vitalik Buterin had an idea to improve Bitcoin \cite{vital0,ether0,ether2,ether3,book0,eth00}. Rather than the block including only transactions, Vitalik wanted them to include sources of code. His motivation was the need to use the blockchain properties to develop decentralized applications, \textbf{(Dapps)} and through creating the world's strongest supercomputer, \cite{ether2}. However,  there was a tall hurdle that Vitalik had to face:  Bitcoin used a scripting language \textbf{Bitcoin Script} \cite{ether2} that wasn't adequate for Vitalik's goal: This language wasn't \textbf{Turing Compete}, namely there are logical problems that one cannot code using it. To achieve Turing completeness, one had to enrich it with loops, a programming feature that \textbf{Bitcoin Script} didn't support.: The developers wanted to prevent tedious efforts such as infinite loops that can significantly slow the network. Vitalik wanted to enrich \textbf{Bitcoin Script} with loops. He didn't get his request. Vitalik moved forward with his idea and established a new blockchain-based network: Ethereum. While Bitcoin acts merely as a crypto-currency,  Ethereum indeed followed Vitalik's vision in two aspects:
\begin{itemize}
\item  It became a platform for developers to create their applications and for users to run these applications \cite{ether0, ether3, smar1}.
\item In contrast to Bitcoin, Ethereum's blocks contain, in addition to transactions, source codes as well. One can run these codes on the network's nodes. 
\end{itemize}
These two achievements allow Ethereum to decentralize many computing tasks and, as mentioned above, develop many \textbf{Dapps}.  We can summarize with the words of Dr. Gavin Wood, one of the Ethereum founders: \textbf{Bitcoin is first and foremost a currency; this is one particular application of a blockchain. However, it is far from the only application. To take a past example of a similar situation, e-mail is a particular use of the internet, and undoubtedly helped popularise it, but there are additional Internet usages} \cite{ether0}.
Ethereum allows more capabilities and requires a more clever environment. In the following sections, we will describe these properties.

\subsection{Ethereum Virtual Machine} 
 We discussed in the previous section the need for improving blockchains with source codes. This improvement brings two threats\cite{vital0, ether2}:
 \begin{itemize}
\item Infinite loops
\item Public accessibility 
\end{itemize}
In this section, we discussed the latter. A blockchain user has the entire chain on his private computer. If we place the sources themselves on the blocks, we may suffer from the following risks:
 \begin{itemize}
\item The source codes can access the private drivers of the users.  
\item If it is plausible to add sources to the blocks, it is easy to deploy viruses
\end{itemize}

To overcome these risks\cite{ether2, ether0}, every participant in Ethereum receives a virtual machine that separates her Ethereum activities from the personal drivers: Ethereum Virtual Machine \textbf{EVM}. The developers upload their sources to the chain, and  \textbf{EVM} executes these sources. The separation from the private drives enhances security.
\subsection{What is Smart Contract?} 
\begin{figure}[h!]
\centering
\includegraphics[width=12cm]{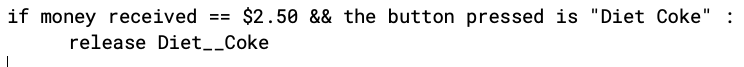}
\caption{ Vending Machine Contract  }
\label{fig01}
\end{figure}

\begin{figure}[h!]
\centering
\includegraphics[width=12cm]{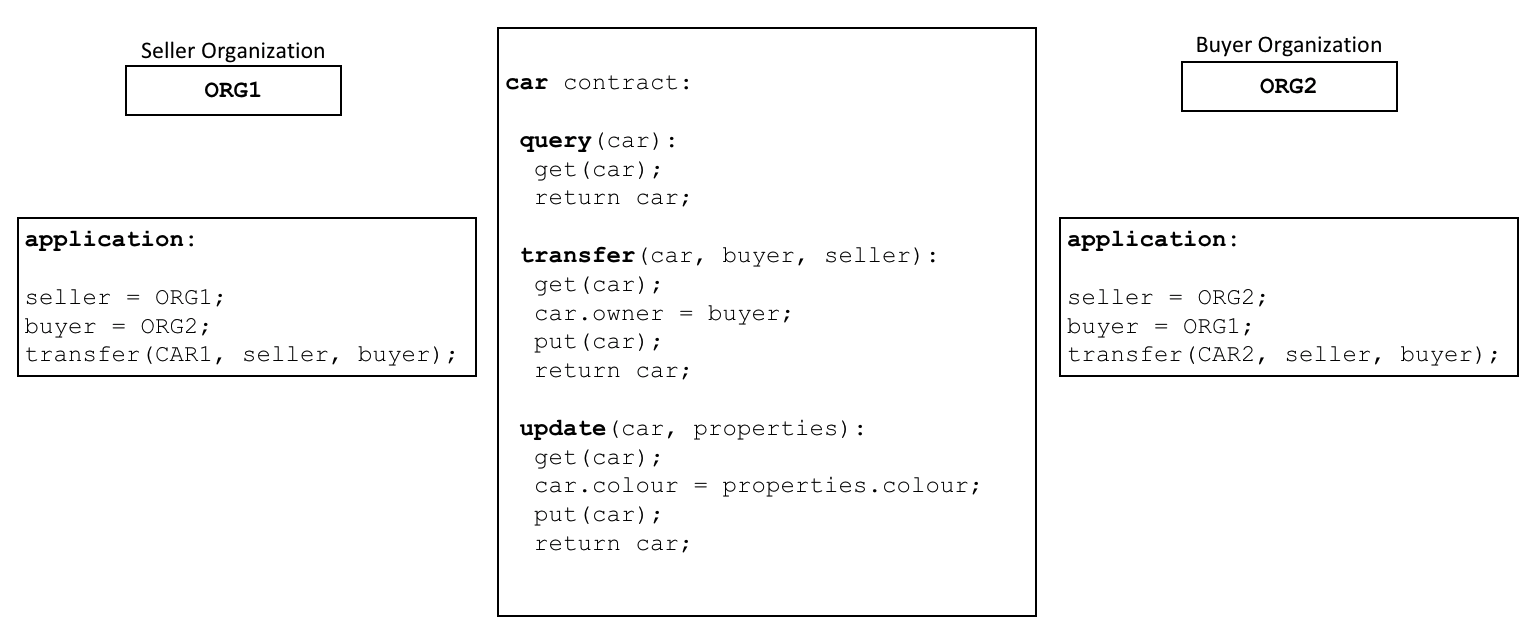}
\caption{ Selling car using a smart contract  }
\label{fig02}
\end{figure}

In a nutshell, smart contracts are merely contracts. The main differences between these contracts and the common "real world" contracts are \cite{vital0, ether01, ether2, smart1,sm22}:
\begin{itemize}
\item Blockchains use them 
\item They are computer programs.
\end{itemize}
The concept of source codes allows \textbf{Dapps} to use them as their back-end for managing the API and the interaction with the blockchain itself. Why do we use the term \textbf{"contract"}?  I will refer to a great post \cite{smart1}, where the writer uses the idea carved on stone contracts. These contracts provide a high level of trust. In a more modern example \cite{smart1}, we can think of a vending machine: It will output a Diet Coke if you provide a certain amount of money and press the button. One can see in fig  \ref{fig01} that this procedure can be programmed \cite{smart1}.
We can create more sophisticated contracts \cite{sm33} as one can see in figure \ref{fig02}

\subsubsection{Smart Contract Compilation}
\begin{figure}[h!]
\centering
\includegraphics[width=12cm]{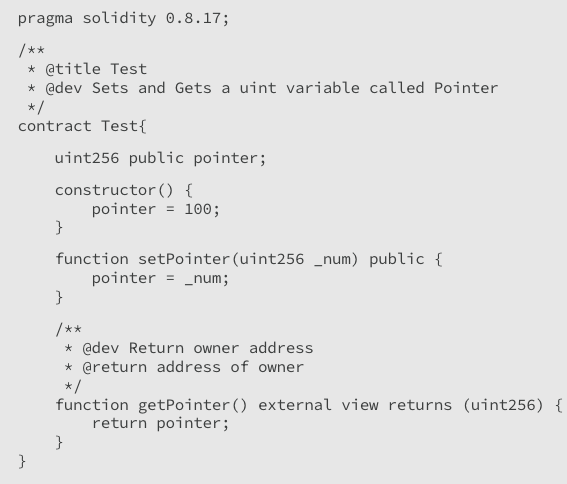}
\caption{ Solidity source example  }
\label{fig03}
\end{figure}

\begin{figure}[h!]
\centering
\includegraphics[width=12cm]{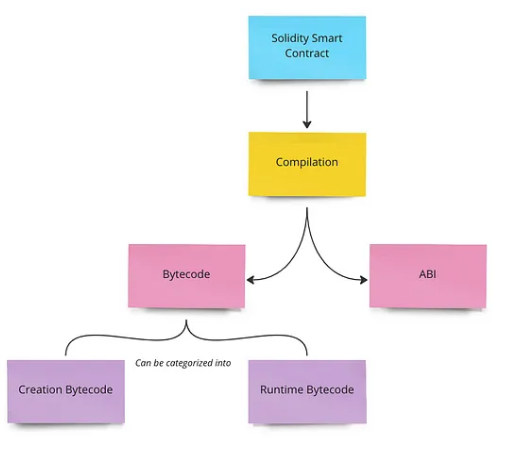}
\caption{ Smart contract's compilation chain  }
\label{fig04}
\end{figure}

\begin{figure}[h!]
\centering
\includegraphics[width=6cm]{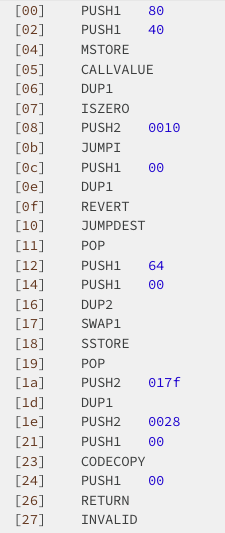}
\caption{ Opcode's Insurrections  }
\label{fig05}
\end{figure}

We use a special language program for smart contracts: \textbf{Solidity}, \cite{book0, ether2}.
An example of a Solidity code \cite{sm0} can be seen in \ref{fig03}, and the entire compilation process\cite{sm0} in \ref{fig04}
We can see that there are two outcomes for the  output of the compilation process :
\begin{itemize} 
\item Application Binary Interface \textbf{ABI} This interface clarifies which functions and which parameters the contract uses \cite{book0, sm0, la}
\item Bytecode - A hexadecimal string that encodes the source code. The developer uploads the bytecode to the EVM.  The EVM  extracts the opcode instructions \cite{book0, sm0}. An example opcode can be seen \cite{sm0} in \ref{fig05} respectively.
\end{itemize}

\subsubsection{Gas and Infinite Loops}
We described the potential problem that Ethereum may encounter with heavy calculations in general and infinite loops in particular. In this section, we will discuss how Ethereum handles this obstacle \cite{vital0, book0, ether2}. Every source code is an ordered collection of commands. Thus, Ethereum uses the concept of \textbf{Gas} \cite{ether2}. The idea is that one has to pay for every computation he wishes to run on the network. Ethereum publishes a pay list that assigns a gas fee for each command.
This approach provides two significant benefits:
\begin{itemize}
\item There are no infinite loops- A user can run a contract only if she can budget the required computational resources
\item Code is written efficiently 
\end{itemize}
Gas is, therefore, a mandatory component for achieving Turing completeness in a blockchain. This concept emphasizes that Ethereum (in contrast to Bitcoin) was developed as a computational platform,  not merely a .crypto-currency.
 

\section{Machine Learning }
\label{ml}
Machine learning \textbf{ML} is not a new discipline. Researchers used tools such as decision trees and probabilistic modeling for decades\cite{andrew1, duda1}. 
The usage of ML has been boosted during the last years by the Deep learning \textbf{DL} revolution that lately reached a highlight with the launch of ChatGPT \cite{chat}. While tools like Random Forest \cite{andrew1, duda1, rf} and XGBoost \cite{xg, xg1, andrew1} are known to most researchers, in this section I will describe some of the deep learning tools that we used. 
\subsection{Generative Pre-trained Transformer-GPT}
\begin{figure}[h!]
\centering
\includegraphics[width=6cm]{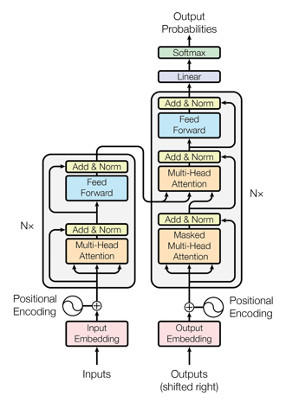}
\caption{ Transformers' Architecture  }
\label{fig09}
\end{figure}

\begin{figure}[h!]
\centering
\includegraphics[width=6cm]{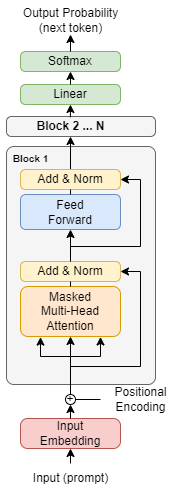}
\caption{ GPT-Decoder Only }
\label{fig10}
\end{figure}

The non-quantitative nature of text challenged the DL world from its inception. Researchers searched for an optimal method to represent text numerically in a way that preserves its intrinsic semantic properties \cite{mik1,atten0}.
In 2017, researchers achieved a breakthrough when they presented a novel architecture:  the \textbf{transformers} \cite{atten1}. Describing transformers in detail is beyond the scope of this paper. However,  one can learn its general structure in \ref{fig09}. The left part of the transformer is called \textbf{"encoder" } and the right \textbf{"decoder"} 
The DL world has focused on transformers since they presented for the first time.In 2018, \textbf{OpenAI} published a new type of transformer: \textbf{GPT} \cite{gpt1}. This architecture is comprised only of the "decoder" part. Hence, it is easier to be trained \ref{fig10}.
As you may assume, this architecture has advanced versions \textbf{GPT2} and \textbf{GPT3.5}. We will discuss the former in the next section, and the latter is the platform for the first version of ChatGPT,

\subsection{Auto-Encoder}
\begin{figure}[h!]
\centering
\includegraphics[width=6cm]{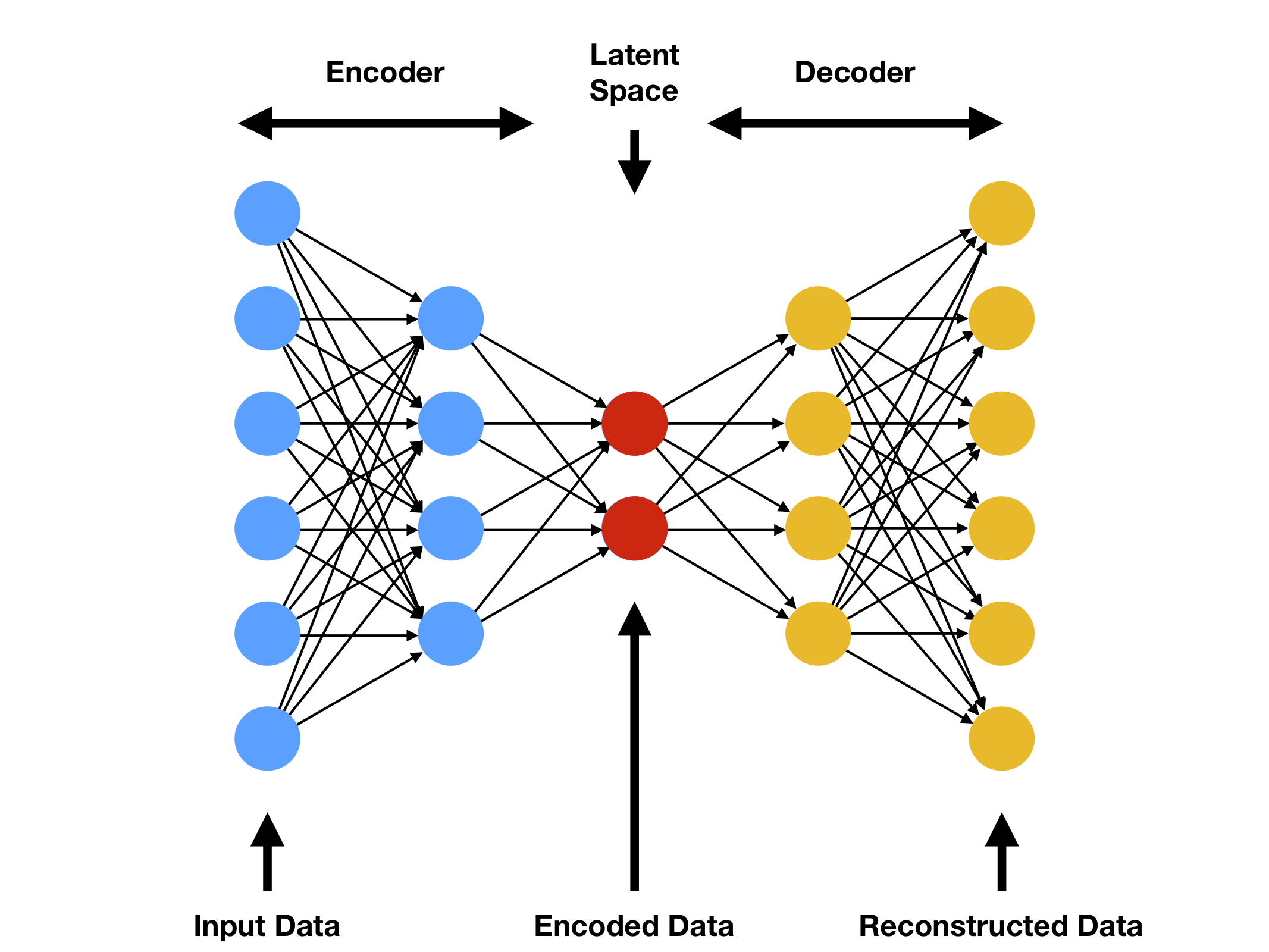}
\caption{ Auto Encoders \cite{ae0}  }
\label{fig11}
\end{figure}

Auto-encoder \cite{andrew1} was one of the first neural network architectures. Most of the neural network architectures focus on tasks such as prediction. Auto-encoder maps the real data into an Euclidean representation as in \ref{fig11}. The auto-encoder consists of two parts:
\begin{itemize}
\item \textbf{Encoder}- Encoder maps the data to a latent space (usually with a lower dimension).
\item \textbf{Decoder}- Decoder maps the latent space back to the raw space  
\end{itemize}
When we complete the training process, the encoding component receives real data as input and outputs an Euclidean representation in a lower dimension.

\subsection{Bayesian Network}
\begin{figure}[h!]
\centering
\includegraphics[width=6cm]{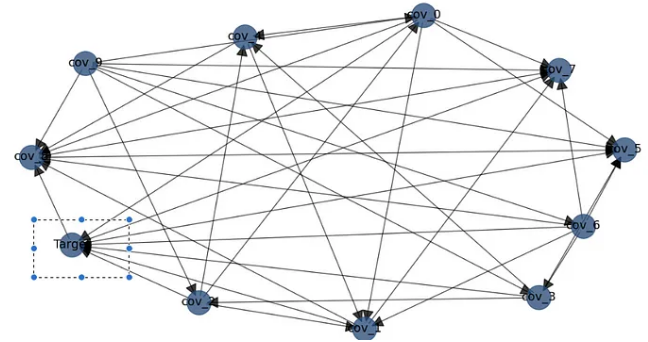}
\caption{ Directed Acyclic Graph - AG  }
\label{fig12}
\end{figure}

Causal inference is a rising research domain \cite{ci0, ci1}. It offers promising results in various disciplines of applied science. Some of its tools can enhance the methodologies by which researchers approach data \cite{ci0, ci1}. A core object in Casual inference is Directed Acyclic Graph \textbf{DAG}, \cite{ci0, ci1}. It represents the actual relations between variables and allows us to perform casual analysis \ref{fig12}.

\begin{figure}[h!]
\centering
\includegraphics[width=6cm]{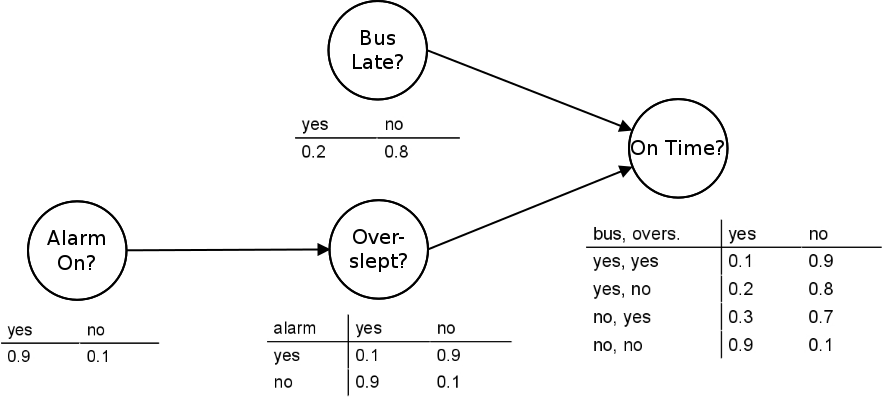}
\caption{ Bayesian Network \cite{bn3} }
\label{fig13}
\end{figure}

However, in real-world data, the arcs' directions in a data graph are often unknown. Researchers use probabilistic tools to approximate these directions. Bayesian networks are such tool \cite{bn0, bn1, bn2}. The idea is that statistical tools will denoise the data and reveal the genuine mutual influence between given variables, as one can see in fig \ref{fig13}. 
In the following sections of this paper, we will describe Ethereum ML research that we did using the described tools for solving some of the problems that we discussed in section \nameref{smart0}

\section{Reentrancy Attack-Real World Example}
\begin{figure}[h!]
\centering
\includegraphics[width=6cm]{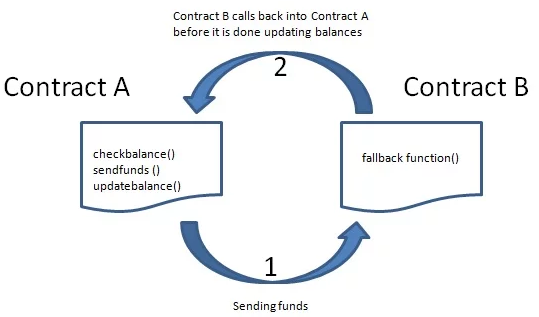}
\caption{ Reentrancy Scheme }
\label{fig06}
\end{figure}

We saw in \nameref{smart0} that indeed smart contracts are a powerful tool. Nevertheless, as with any other technology it comes with its "novel" risks. Smart contracts are used in Ethereum transactions, hence its vulnerabilities can be exploited in these transactions. Since it is merely a source code, we can suspect that if we don't develop it properly, we may suffer drawbacks \cite{ether3, re0, re1,re2, book0}. One of the most common attacks of smart contracts is the \textbf{reentrancy attack}. In such an attack, a fund transaction takes place. Rather than updating the paying account and transferring the fund in the following step, the contract transfers the fund before. As a result, one can call the transfer function infinitely many times and transfer the entire money of the victim. A graphical scenario of \textbf{reentrancy attack} is presented in fig \ref{fig06}. This scheme is presented in \cite{re0, re1}

\begin{figure}[h!]
\centering
\includegraphics[width=6cm]{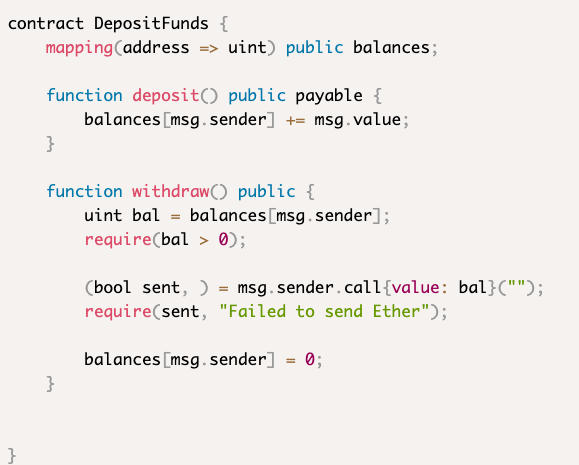}
\caption{ Deposit Function }
\label{fig07}
\end{figure}

\begin{figure}[h!]
\centering
\includegraphics[width=6cm]{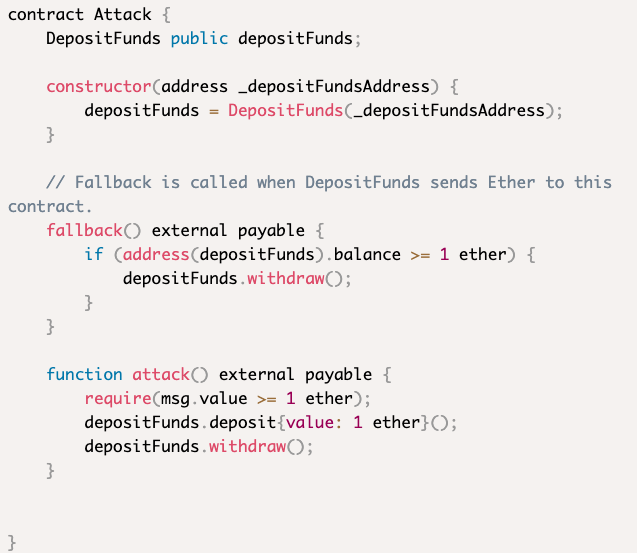}
\caption{ Reentrancy Attack function }
\label{fig08}
\end{figure}

We present a Solidity implementation of the reentrancy attack in figures \ref{fig07} and \ref{fig08} \cite{re0}
The most famous  \textbf{reentrancy attack} is the \textbf{DAO Attack}. It took place in 2017. 60 million dollars were stolen. Moreover, the main outcome of this attack was a hard fork of Ethereum, that led to two communities. The readers can learn about this attack and its impacts on Ethereum \cite{vital0, ether2,ether3}.
There were some additional attacks after the DAO such as \cite{re0} :
\begin{itemize}
\item  Uniswap/Lendf.Me hacks (April 2020) –  25 Million dollars, attacked by a hacker using reentrancy.
\item The BurgerSwap hack (May 2021) – 7.2 Million dollars because of a  and a reentrancy exploit.
\end{itemize}

\section{Research Methodology}
We will present three ML types of research that focus on smart contracts:  
\begin{itemize}
\item Smart contract detection based on the opcode. This model  relies on the GPT2 model \cite{gpt1,gpt2, hf} 
\item Smart contract detection based on the opcode. This model relies on CodeLlama fine-tuning \cite{llama, llama1, codel1}
\item Malicious transaction detection which is based mainly on gas-driven variables.
\end{itemize}

In these studies, we develop ML binary models using common tools \nameref{ml} or \cite{gpt1, gpt2, llama, codel1}. In \nameref{trx0}, we describe in detail the transactions model.
We assume that two main manners characterize malicious transactions:
\begin{itemize}
\item The user is willing to pay a high amount of gas to place the transaction as quickly as possible
\item The user may choose different approaches to use common smart contract functions 
\end{itemize}
These two assumptions gave us the guidelines for choosing the relevant variables to our model. 
In the smart contract models, we followed the code generation approach \cite{gpt1, codel1}and used the code LLM models. For the solidity study, we "innovatively" fine-tuned a CodeLlama binary model \cite{nkl}. 
\subsection {Malicious Data}
When we develop binary models we need labeled data for both populations (e.g. benign and malicious). While benign data is always available (as an example see \cite{forta}), there are some obstacles in aggregating malicious data:  
\begin{itemize}
\item Both Solidity sources and ABI of malicious contracts are seldom available
\item It is difficult to label contact as malicious unless it is already harmful. 
\item Transactions are rarely labeled as malicious 
\end{itemize}
We, therefore, used common websites such as \cite{cha, defi} that indicate exploiting contracts or scam transactions. It allowed us to collect "bad reputation" contracts and indicate them as malicious

\section{The Opcode project}
\label{opc}
\subsection{Smart Contracts' Detection}
\begin{figure}[h!]
\centering
\includegraphics[width=6cm]{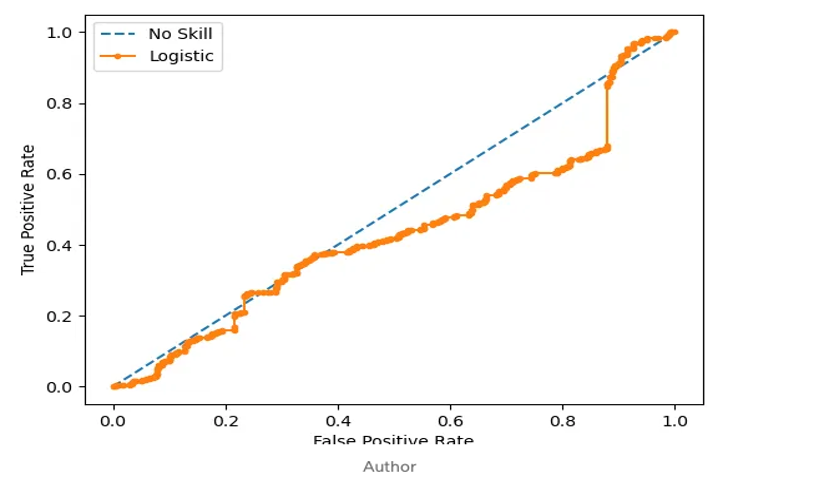}
\caption{ GPT2 False Positive Vs. recall }
\label{fig14}
\end{figure}

\begin{figure}[h!]
\centering
\includegraphics[width=6cm]{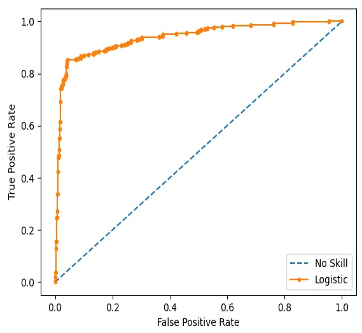}
\caption{ Trained GPT2 False Positive Vs. recall }
\label{fig15}
\end{figure}

The idea of using an opcode source for detecting malicious smart contracts is not novel \cite{forta}. Opcode is  expected to be a beneficial prediction feature due to several reasons:
\begin{itemize}
\item The code is easy to disassemble from the bytecode
\item The commands are human-readable, which allows NLP tools usage. 
\item We can detect statistical patterns on verified and non-verified contracts  \cite{forta}.
\end{itemize}
We aspire to develop a classification model that receives opcode sources of smart contracts and detects whether they are malicious. We decided to use GPT2 \cite{gpt2} that is provided by Huggingface \cite{hf}. The motivation to use this model is that it simply allows us to consider each contract source as a textual section where each command is a word. Our data included 10k benign smart contracts and about 800 malicious ones. Namely, our data is extremely imbalanced. Before starting the training, we measure the performances of GPT2 as presented in \ref{fig14}
One can see that GPT2 performs nearly as random. We use GPT2's weights as our initial weights and train its architecture using the data that we have described. In early trials, our naive approach was flattening each contract and providing it to the model. The results weren't good, which led us to perform some trial and error for improvements. Truncating the prefix and restricting the length to 600 commands (with padding if needed) provided optimal results, as shown in fig \ref{fig15}.
 
\subsection{Opcode Graphs}
\begin{figure}[!tbp]
  \centering
  \subfloat[Malicious DAG]{\includegraphics[width=0.65\textwidth]{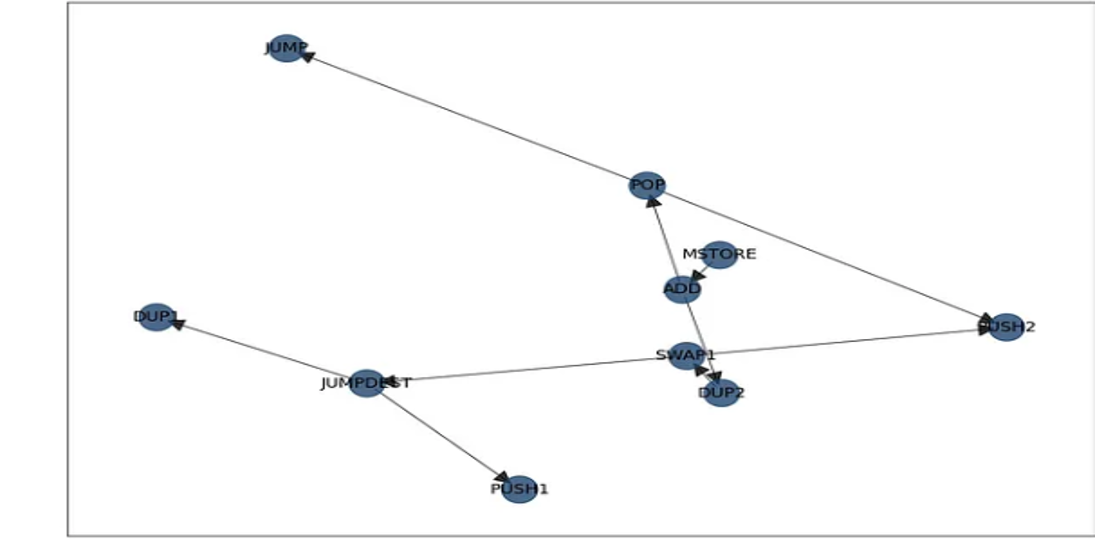}\label{fig:f1}}
  \hfill
  \subfloat[Benign DAG]{\includegraphics[width=0.65\textwidth]{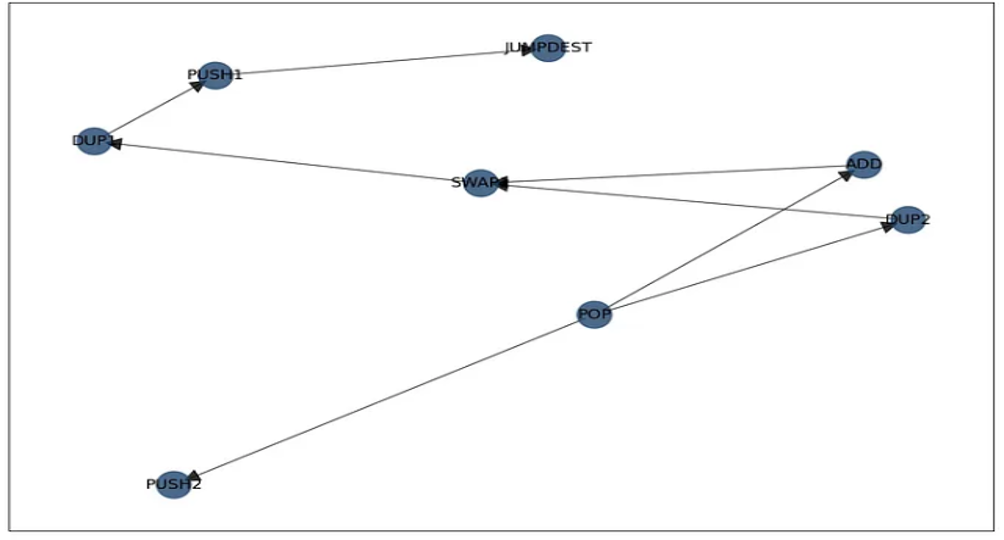}\label{fig:f2}}
  \caption{DAGs Comparison}
  \label{fig17}
\end{figure}

As in nearly every other ML project in cyber, prediction is insufficient: one needs to demystify which features are cardinal in determining the model's decision. In our project, there are no features besides the opcode commands. Thus, we can focus only on studying the command occurrences and their inner relations, such as mutual occurrence (or anti-occurrence). We used Python's bnlearn \cite{bn1} to generate the DAGs of the benign and the malicious contracts. We present the comparison between the two DAGs in figure  \ref{fig17}

The main objective of this comparison is to show that a difference between the two DAGs, particularly their arcs, exists.

\section{The Solidity Project }
\label{sol0}
The success of the Opcode project motivated us to develop another prediction model: a Solidity model. After a short research, we realized that:
\begin{itemize}
    \item There are no available databases of malicious contracts. 
    \item A malicious contract is often unverified -we don't have the source.
\end{itemize}
The latter requires a remark about Ethereum manner: In contrast to bytecode, we don't upload Solidity sources to the blockchain. When we deploy a contract, we don't have to verify it, which leads the source to remain unobservable.  This manner is counter "blockchain": we would expect that in an environment that aspires to be social, the community has to verify a code and not the developer.
We needed both to increase our malicious data compared to the opcode project and generate source code. We achieved the former using some websites such as Chainabuse \cite{cha}  and the latter using a web decompiler \cite{dec00}.
Our data includes about 100k benign contracts and about 2000 malicious contracts with their (recompiled) Solidity sources. We used the Huggingface database as our source of benign contracts \cite{hfdb}.  

\subsection{CodeLlama}
We wish to use the data to train a model. Since we are in the large language models \textbf{LLM} era, a natural step is searching for such a model. Llama is a collection of text generation LLMs that Meta trained \cite{llama, llama1}, and includes code generative models \textbf{CodeLlama}. For training these models, Meta used Java and Python sources \cite{codel1} for training these models, which can fit our needs. We need to fit them for classification tasks, which requires the following:
\begin{itemize}
    \item Modifying the architecture
    \item Perform a model fine-tuning  
\end{itemize}
The second clause is mandatory since the training process of the basic model included different programming languages. We wish to focus solely on Solidity and its malicious-benign differences. The solution of the first clause is described in detail in \cite{nkl}. We managed to achieve results of around 70 percent accuracy. However, it is a 7B model with very little data. Since we cannot increase our data massively (we don't have sufficient malicious contracts), we searched for a smaller model.

\subsection{The Solidity Generator}
\begin{figure}[h!]
\centering
\includegraphics[width=6cm]{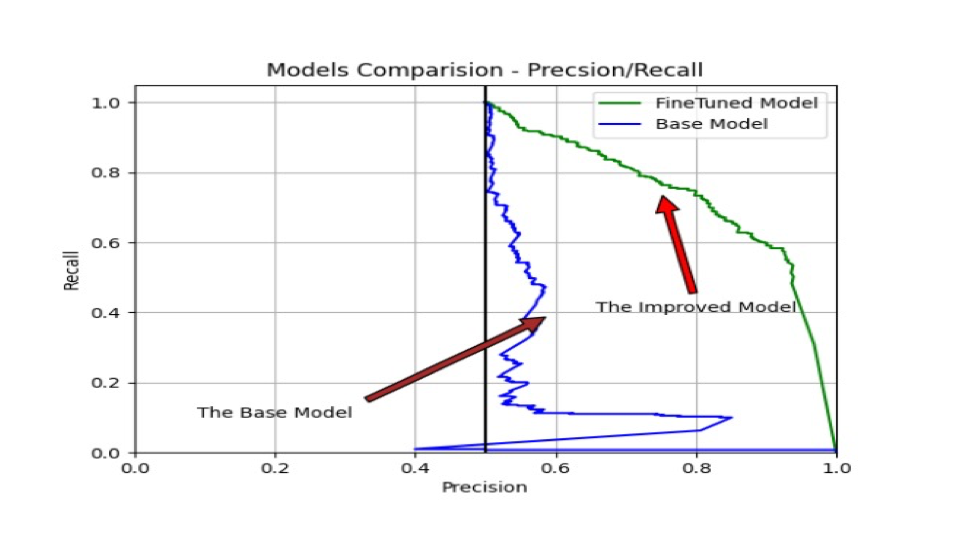}
\caption{ The Solidity Classification Model }
\label{fig19}
\end{figure}

While making a quick search on the web, we found a solidity generator model \cite{solg}, which is based on \textbf{BERT} \cite{bert} architecture. We used a similar fashion to \cite{nkl} to modify it. However, this model is significantly smaller (about 500M). Thus, we can better both validate its results and deploy it. We present the results in fig \ref{fig19} 
Before we analyze the results, we describe the axes KPI's 
\begin{itemize}
    \item X-axes: Precision- The probability that a detected contract is malicious.
    \item Y-axes: Recall - The probability of detecting a malicious contract. 
\end{itemize}
The base model is the model before we fine-tuned it. On one hand, the results are around 80 percent, which is an outcome of the absence of data. On the other hand, our fine-tuned model is significantly better than the base model. This indicates that the data is "trainable". We have a good reason to assume that enriching the data, particularly with malicious sources, will provide a better model.

\section{ML Study of Malicious Transactions}
\label{trx0}
In this section, we discuss the ML study of the transactions. We recall that we restrict ourselves to transactions in which the receiver is a smart contract. We will base our model features on data, that we get from the web3 package \cite{web3}.
\subsection{Data}
We must aggregate malicious and benign transactions to train a malicious detection engine. For aggregating malicious transactions we used websites such as \textbf{de-fi} and \textbf{chainabuse} \cite{chabus, defi} as well as information about malicious smart contracts that one can easily find \cite{chabus}.
We aggregated about 15K malicious transactions. For aggregating benign transactions we used two methods:
\begin{itemize}
    \item Sampled "modern" transactions (namely from the very last blocks.
    \item  Sampled from blocks we took malicious transactions to overcome potential data temporality. 
\end{itemize}
For generating features we used the structures of \textbf{gettransaction} and \textbf{gettransactionreceipt} of  Python's web3 \cite{web3}.
\subsubsection{Our Features}
The web3 package provides a massive amount of data about the Ethereum transactions. Not all of them provide information that is beneficial for malicious detection. Our research focused on variables that describe aspects of gas bid-ask, such as the value, and (as we present in the next subsection ) the input's functions. 
We briefly describe some of the variables:
\begin{itemize}
    \item \textbf{effectiveGasPrice} -The effective gas price that was paid
    \item  \textbf{cumulativeGasUsed} - The amount of gas in the block until the relevant transaction,
    \item  \textbf{to} - Obvious field . Here we need to verify that it is a contract.
    \item \textbf{ logs} The number of logs that appeared during the execution of the smart contract.
    \item  \textbf {gasUsed/gasPrice} The gas was used in the transaction and its cost in wei.
    \item  \textbf{type} It indicates the mechanism of Ethereum that reflects the gs pricing, in particular, it reflects if the transaction follows EIP-1559 or not.
    \item  \textbf{gas} -Maximum amount of gas that the sender is willing to pay
     \item  \textbf{maxFeePerGas}- Amount of gas that one is willing to pay for computational resources
    \item  \textbf{ maxPriorityFeePerGas}- Maximum tip one is willing to pay
\end{itemize}
We can generate ML features from these variables. These features are the input of an ML model. The motivation to focus on gas consumption, gas price, and gas fee,  is derived from the assumption that a potential culprit will offer a high amount of gas to execute his transactions.  
\subsection{The Model}
Since we are using tabular data, the natural approach is using XGBOOST \cite{xg, xg1}
The input features we provided XGBOOST are the variables we discussed in the previous section with some additional aggregators. We trained the XGBOOST model and got the following ROC:
\begin{figure}[h!]
\centering
\includegraphics[width=6cm]{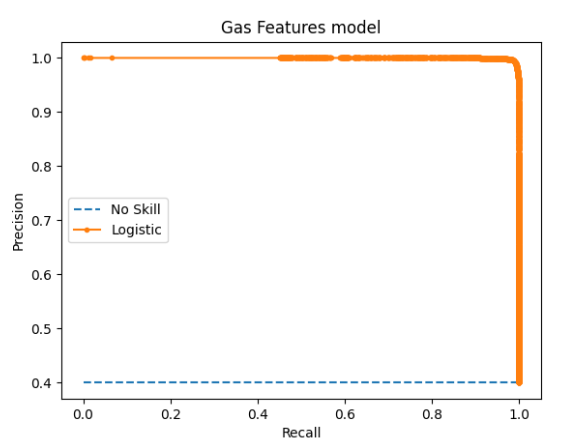}
\caption{ ROC curve of the xgboost model }
\label{gas}
\end{figure}
In addition, we present the feature importance of the xgboost model:
\begin{figure}[h!]
\centering
\includegraphics[width=6cm]{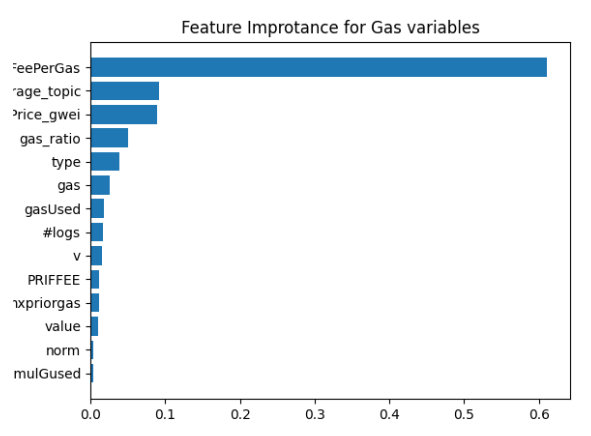}
\caption{ Feature Importance of our model }
\label{gas}
\end{figure}
We can see that the gas variables are the most significant as one may expect. 

\subsection{The 4bytes Analysis}
Every transaction includes a field called \textbf{input} which is a hexadecimal signature that indicates the functions of the smart contract and their inputs. There is a website \textbf{4bytes} \cite{4bytes} that provides a map from many of these hexadecimal signatures to their actual smart contracts' functions. Our next step in the research is using this map to generate statistics about such functions. We developed a mapping from the Hexa signatures to the functions. We denote the latter \textbf{hexdic}. The following graph represents the distribution of signatures:
\begin{figure}[h!]
\centering
\includegraphics[width=6cm]{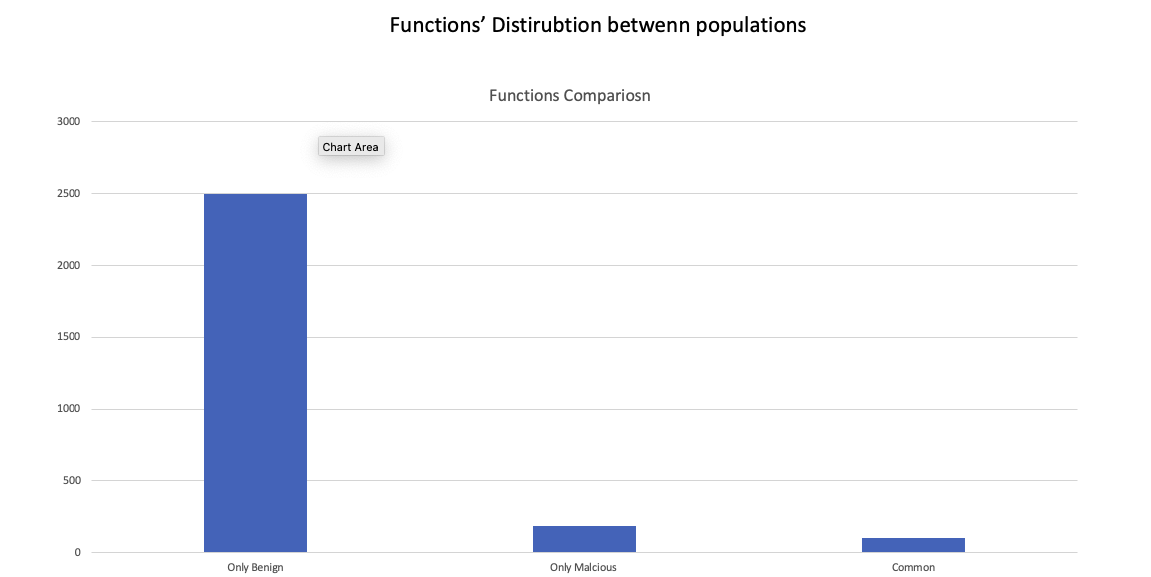}
\caption{ Signatures' histogram }
\label{hist}
\end{figure}
One can see that most of the signatures that we can map appear only in benign transactions. on the other hand, we have about 60k benign transitions and about 100 signatures appear only in the malicious which is a potential indicator.
The next KPI that we studied was the length of each input. The results are presented
\begin{figure}[h!]
\centering
\includegraphics[width=6cm]{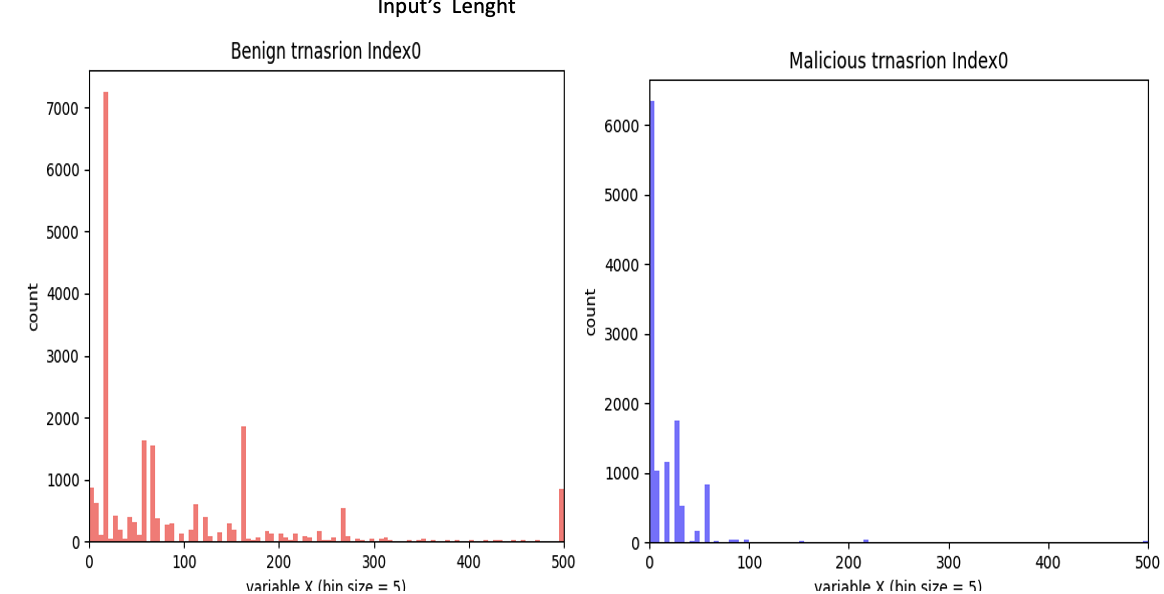}
\caption{ Signatures' Length Distribution }
\label{len}
\end{figure}
One can see differences between the distribution lengths which we may use for detection using a Bayesian or a frequentist fashion. The following graph represents the distribution of the valid amount of signatures (valid -we have in our map)

\begin{figure}[!tbp]
  \centering
  \subfloat[Valid Sig]{\includegraphics[width=0.65\textwidth]{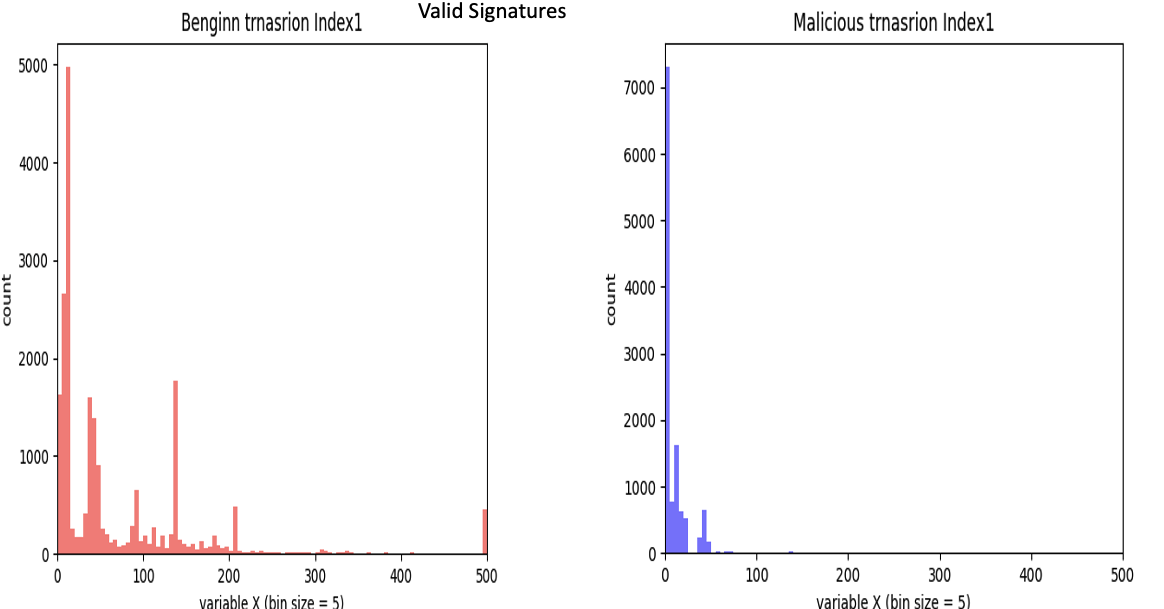}\label{fig:f1}}
  \hfill
  \subfloat[Prop valid ]{\includegraphics[width=0.65\textwidth]{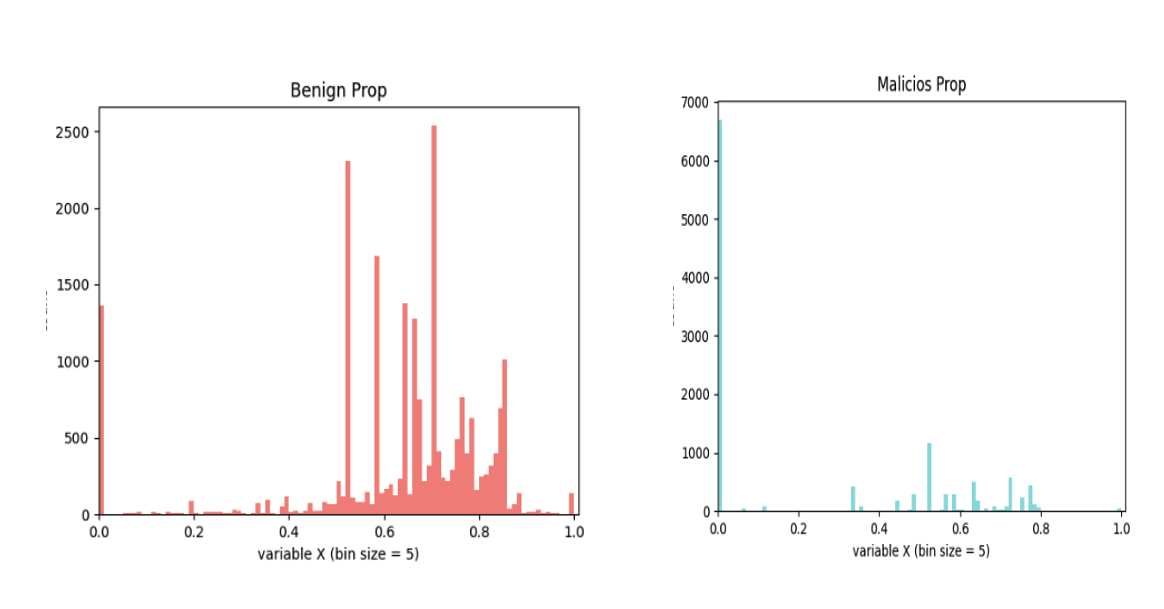}\label{fig:f2}}
  \caption{Valid signature Distribution}
  \label{valid}
\end{figure}
To provide some expert knowledge, we present a list of functions, that are common for both being and malicious 
\begin{figure}[!tbp]
  \centering
  \subfloat[Malicious Functions  ]{\includegraphics[width=0.65\textwidth]{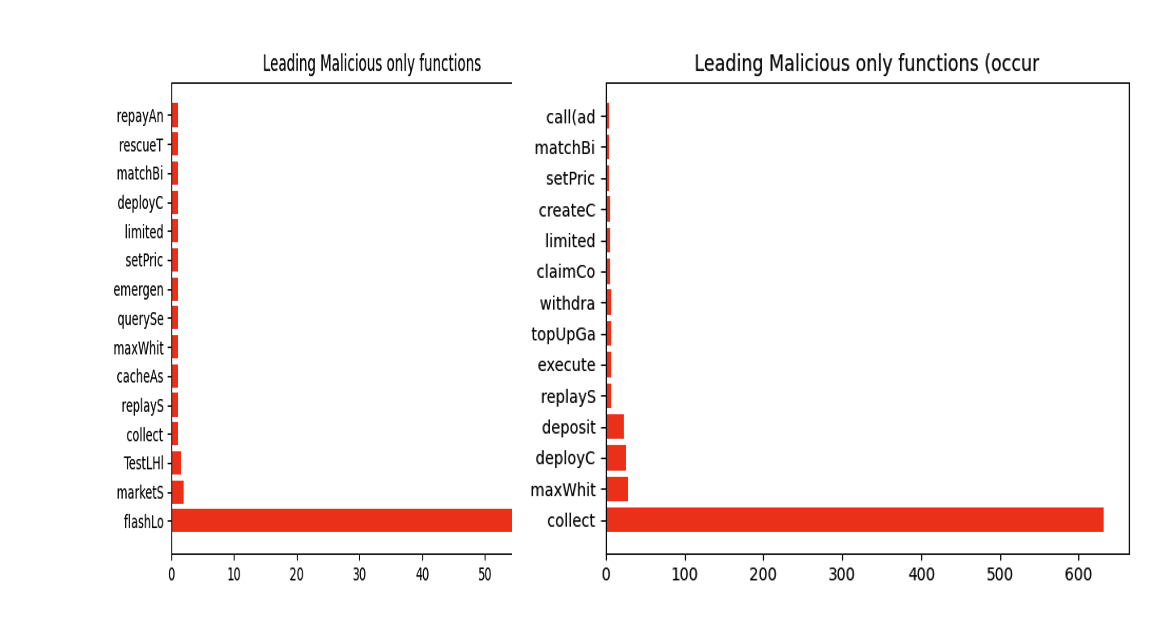}\label{fig:f1}}
  \hfill
  \subfloat[Benign  Functions ]{\includegraphics[width=0.65\textwidth]{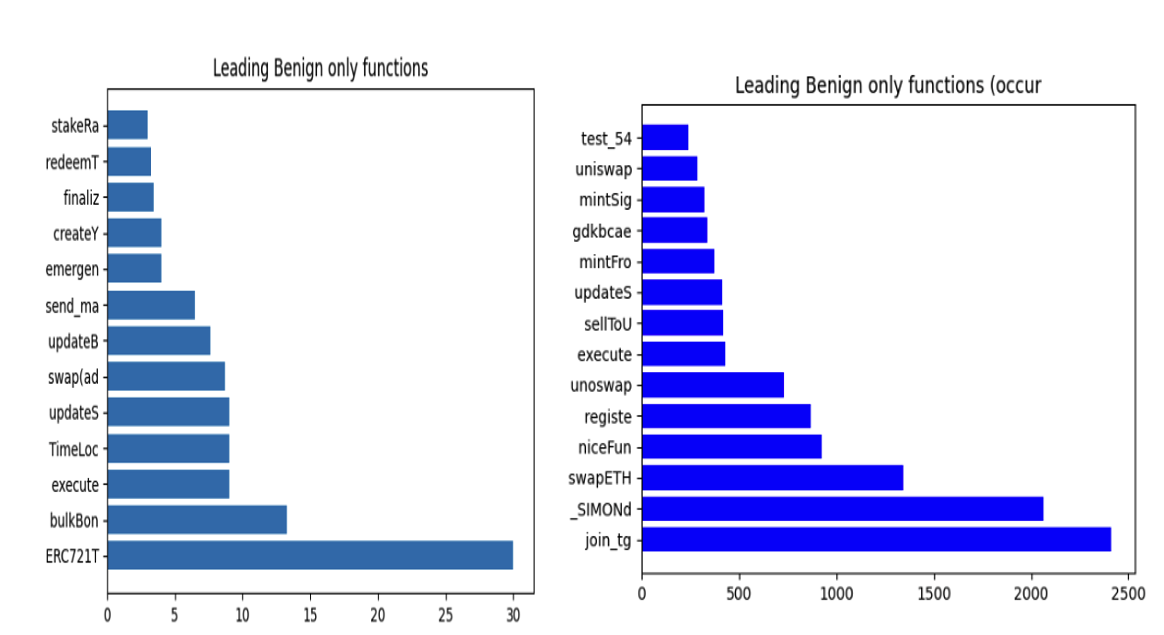}\label{fig:f2}}
  \caption{Valid signature Distribution}
  \label{funcs}
\end{figure}
Finally, we wish to add this information to the model. The following source presented one of our approaches
\begin{lstlisting}
    def generate_octet_features(self,tx_input ):
        list_of_ordered_octets = textwrap.fill(tx_input[2:], 8).split()
        n_octets =len(list_of_ordered_octets)
        valid_octet=0
        mal_octet=0
        benign_octet=0
        for octet in list_of_ordered_octets:
            mm = '0x' + octet
            if '0x' + octet in self.hex_dic:
                valid_octet += 1

                if self.hex_dic[mm] in self.mal_dic:
                   mal_octet+=1
                   continue
                if self.hex_dic[mm] in self.benign_dic:
                    benign_octet+=1
                    continue
        return n_octets, valid_octet, benign_octet,mal_octet
\end{lstlisting}
The model results are presented in the following ROC:
\begin{figure}[h!]
\centering
\includegraphics[width=6cm]{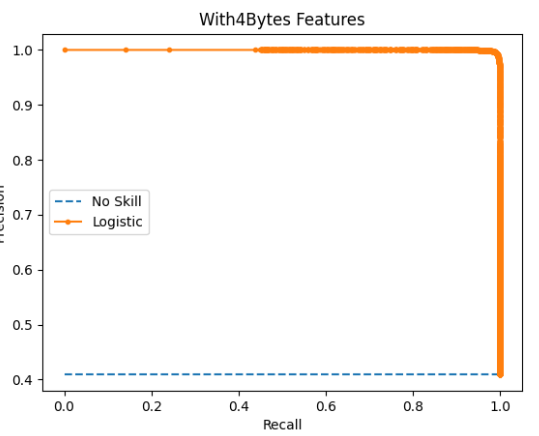}
\caption{Model's ROC with 4bytes }
\label{len11}
\end{figure}
We add the feature importance as well:
\begin{figure}[h!]
\centering
\includegraphics[width=6cm]{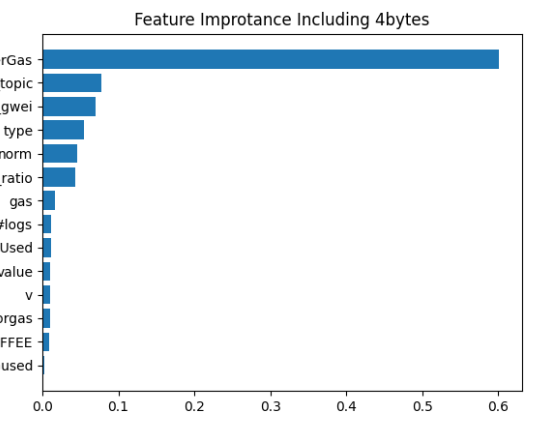}
\caption{Model's Features' importance  with 4bytes }
\label{len21}
\end{figure}
It can be seen that the results are slightly better but not in a significant way. We can explain it by the fact that many transactions are empty. We can deduce that the 4bytes method can be beneficial as an anomaly tool that assists xgboost. 
\subsection{DL Early Study}
in many scenarios, it is difficult to have malicious transitions. We train an auto-encoder only with benign transactions to detect anomalies.   We projected the representation, on a PCA space \cite{pca}, as shown in figure \ref{fig22} : 

\begin{figure}[!tbp]
  \centering
  \subfloat[Components 0 1 2]{\includegraphics[width=0.75\textwidth]{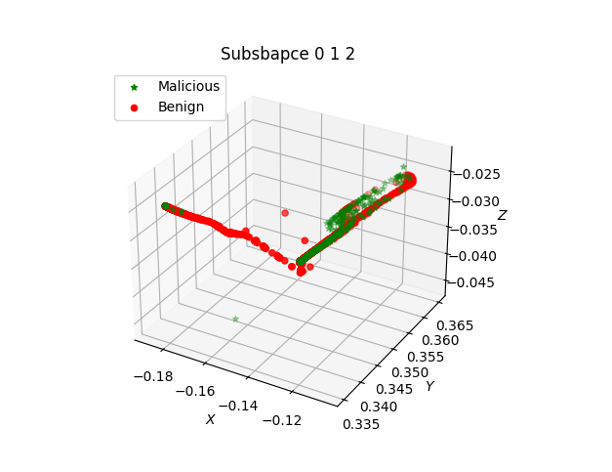}\label{fig:fa}}
  \hfill
  \subfloat[Components 2 3 4]{\includegraphics[width=0.65\textwidth]{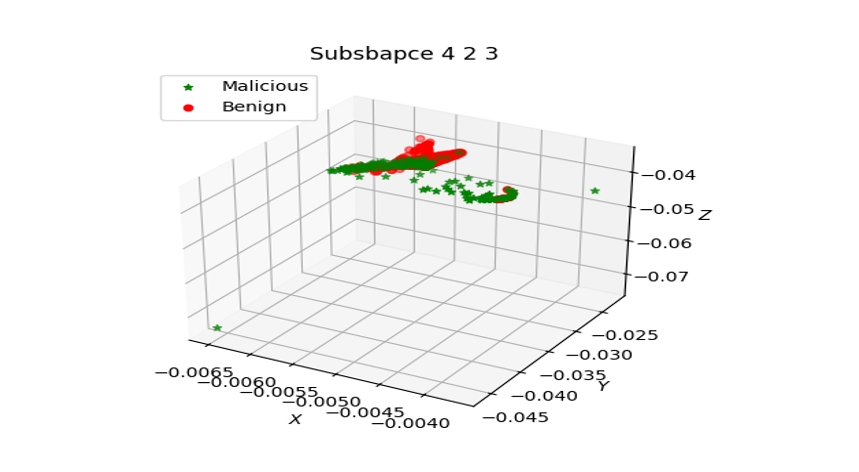}\label{fig:fb}}
   \hfill
  \subfloat[Components 0 1 4]{\includegraphics[width=0.65\textwidth]{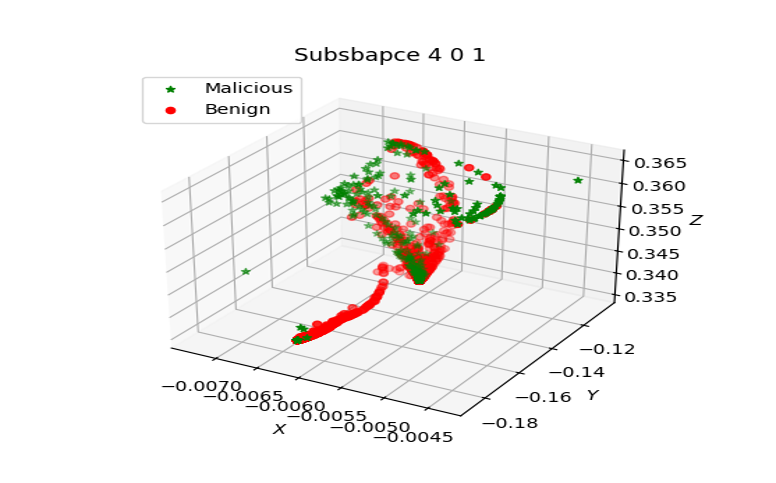}\label{fig:fc}}
   
  \caption{Transactions Projected on PCA}
  \label{fig22}
\end{figure}
One can see that we achieve a separation between the malicious transactions and the benign ones. The graph doesn't indicate how well the separation is, but it suggests that DL methods can be beneficial in detecting differences between the two transaction populations.

\section{Summary and Future Steps}
We have presented several projects that use ML to handle maliciousness in Ethereum. This tool seems to be beneficial on various levels. Following our work, the next natural steps need to focus on two directions:
\begin{itemize}
    \item Improving the data collection methodology. In particular, malicious contracts.
    \item Finding novel methods to exploit LLMs to solve Web3 tasks in general and blockchain security in particular,
    \item Improving the  transactions detection models
    \item Perform a better usage of the 4bytes information (using anomaly detection and probabilistic methods)
\end{itemize}


%
\end{document}